\documentclass[english,floatfix,english,superscriptaddress,notitlepage,longbibliography]{revtex4-1}
\usepackage[T1]{fontenc}
\usepackage[utf8]{inputenc}
\usepackage{xcolor}
\usepackage{pdfcolmk}
\usepackage{amsmath}
\usepackage{amssymb}
\usepackage{graphicx}
\PassOptionsToPackage{normalem}{ulem}
\usepackage{ulem}

\makeatletter

\providecolor{lyxadded}{rgb}{0,0,1}
\providecolor{lyxdeleted}{rgb}{1,0,0}

\DeclareRobustCommand{\lyxsout}[1]{\ifx\\#1\else\sout{#1}\fi}

\usepackage{xcolor}
\definecolor{darkblue}{rgb}{0.1,0.2,0.6} 
\definecolor{lightblue}{rgb}{0.1,0.1,1.0}
\definecolor{darkred}{rgb}{0.8,0.1,0.2}
\usepackage[colorlinks,citecolor=lightblue,linkcolor=darkblue,urlcolor=lightblue] {hyperref}
\renewcommand{\BibitemShut}[1]{}

\makeatother

\usepackage{babel}
\begin{document}
\global\long\def\E{\mathrm{e}}%
\global\long\def\D{\mathrm{d}}%
\global\long\def\I{\mathrm{i}}%
\global\long\def\mat#1{\mathsf{#1}}%
\global\long\def\vec#1{\mathsf{#1}}%
\global\long\def\cf{\textit{cf.}}%
\global\long\def\ie{\textit{i.e.}}%
\global\long\def\eg{\textit{e.g.}}%
\global\long\def\vs{\textit{vs.}}%
\global\long\def\ket#1{\left|#1\right\rangle }%
\global\long\def\etal{\textit{et al.}}%
\global\long\def\tr{\text{Tr}\,}%
\global\long\def\im{\text{Im}\,}%
\global\long\def\re{\text{Re}\,}%
\global\long\def\bra#1{\left\langle #1\right|}%
\global\long\def\braket#1#2{\left.\left\langle #1\right|#2\right\rangle }%
\global\long\def\obracket#1#2#3{\left\langle #1\right|#2\left|#3\right\rangle }%
\global\long\def\proj#1#2{\left.\left.\left|#1\right\rangle \right\langle #2\right|}%
\global\long\def\N{\mathcal{N}}%

\title{Multifractality and its role in anomalous transport in the disordered
XXZ spin-chain}
\author{David J. Luitz}
\email{dluitz@pks.mpg.de}

\affiliation{Max-Planck-Institut für Physik komplexer Systeme, Nöthnitzer Str. 38,
01187 Dresden, Germany}
\author{Ivan M. Khaymovich}
\email{ivan.khaymovich@pks.mpg.de}

\affiliation{Max-Planck-Institut für Physik komplexer Systeme, Nöthnitzer Str. 38,
01187 Dresden, Germany}
\author{Yevgeny Bar Lev}
\email{ybarlev@bgu.ac.il}

\affiliation{Department of Physics, Ben-Gurion University of the Negev, Beer-Sheva
84105, Israel}
\begin{abstract}
The disordered XXZ model is a prototype model of the many-body localization
transition (MBL). Despite numerous studies of this model, the available
numerical evidence of multifractality of its eigenstates is not very
conclusive due severe finite size effects. Moreover it is not clear
if similarly to the case of single-particle physics, multifractal
properties of the many-body eigenstates are related to anomalous transport,
which is observed in this model. In this work, using a state-of-the-art,
massively parallel, numerically exact method, we study systems of
up to 24 spins and show that a large fraction of the delocalized phase
flows towards ergodicity in the thermodynamic limit, while a region
immediately preceding the MBL transition appears to be multifractal
in this limit. We discuss the implication of our finding on the mechanism
of subdiffusive transport.
\end{abstract}
\maketitle

\section{Introduction}

Metal-insulator transitions are central in condensed matter physics.
In most of these transitions the insulating phase is gapped and the
conductivity is mediated by thermal activation across the gap. It
is thus exponentially suppressed at sufficiently low temperatures
strictly vanishing at absolute zero. However in the presence of strong
quenched disorder, and in the absence of interactions, a different
kind of metal-insulator transition is possible, which is called the
Anderson localization transition \citep{Anderson1958b}. Across the
Anderson transition the spectrum is gapless and the transition occurs
due to the change in the nature of the eigenfunctions \footnote{The transition occurs only at three dimensions or higher for tight-binding
models without spin-orbit coupling, and in any dimension for long-range
random matrix models as discussed, e.g., in Refs.~\citep{Levitov1989,Levitov1990,Evers2008a,fyodorov2009anderson,Yuzbashyan_JPhysA2009_Exact_solution,Nosov2019correlation}} \citep{Anderson1958b}. In the metallic phase the eigenfunctions
are ergodic and extended, namely the probability to find a particle
in a certain position is approximately uniform in space. On the other
hand, on the insulating part the eigenfunctions are localized, such
that the particle is found in the vicinity of a certain point. The
Anderson transition point is special, since at this point the eigenfunctions
are neither ergodic and extended nor localized; they cover a sub-extensive
number of sites, a situation which is called multifractality or nonergodic
extended phase \citep{Evers2008a}. The spatial structure of the eigenfunctions
is directly related to the transport in the system. Ergodic and extended
eigenfunctions yield diffusion, while localized eigenfunctions suppress
all transport all together. At the critical point, the system is known
to have subdiffusive transport, with a fixed dynamical exponent \citep{Evers2008a}.

Almost 15 years ago, it was shown that sufficiently weak interactions
between the particles do not destroy the Anderson insulator, but induce
a transition, known as the many-body localization (MBL) transition
between an delocalized and localized phases \citep{Basko2006a} (see
\citep{Abanin2017} for a recent review). Signatures of MBL were observed
in ultracold atomic gases on optical lattices both in one-dimensional
\citep{Schreiber2015a,Bordia2015,Smith2015} and two-dimensional systems
\citep{Choi2016}. Similarly to the Anderson transition, the MBL transition
was believed to be a \emph{finite} temperature transition between
a diffusive metal and an insulator, with the crucial difference that
in the insulating phase, the conductivity of a thermodynamically large
system is \emph{strictly} zero even at finite temperature \citep{Basko2006a,Gornyi2005}.
A number of numerical studies demonstrated later, that for one-dimensional
systems with bounded energy density, transport in the delocalized
phase is \emph{subdiffusive}, and thus conductivity in the thermodynamic
limit vanishes through the entire phase diagram \citep{BarLev2014,Lev2014,Agarwal2014,Luitz2015a,Znidaric2016}.
In addition to the anomalous transport, the delocalized phase shows
sublinear growth of entanglement entropy \citep{Vosk2014,Potter2015,Luitz2015a},
suppressed spreading of entanglement \citep{Luitz2017,Lezama2018,lezama_power-law_2019},
intermediate statistics of eigenvalue spacing \citep{Serbyn2015}
and satisfies only a modified version of the eigenstates thermalization
hypothesis (ETH) \citep{Luitz2016b} (see \citep{Luitz2016c} for
a detailed review of the properties of the delocalized phase). A phenomenological
explanation of the anomalous dynamical properties of the delocalized
phase, based on rare blocking regions, was provided in Refs.~\citep{Agarwal2014,Gopalakrishnan2015a}
(see also recent review \citep{Agarwal2016_review}), however a number
of predictions of this theory are not entirely consistent with numerical
studies \citep{BarLev2015,BarLev2017a,mace_fibonacci_2019} and experiments
\citep{Luschen2016a} (although there is also supporting numerical
evidence \citep{Znidaric2018a}).

Since anomalous relaxation and transport are in many cases related
to multifractality of the eigenstates \citep{Ketzmerick1997,Ohtsuki1997},
a natural question to ask is whether a similar relation exists also
for the delocalized phase in systems which exhibit the MBL transition.
This direction of thought is evermore suggestive, since MBL is often
viewed as Anderson localization in Fock space, or more concretely
on a complicated high-dimensional graph, where the nodes are Fock
states, and the connectivity between them is mediated by the Hamiltonian
(cf. discussion in \citep{Alet2017}). Since the structure of this
graph is rather involved it is normally approximated by either the
Bethe lattice \citep{Altshuler1997} or random-regular graphs (RRG,
see also review by Imbrie \emph{et al. }\citep{Imbrie2016a}). In
addition, the disorder residing on the nodes of this graph, is highly
correlated, a feature which was shown to be important for MBL \citep{Ghosh2019a,Roy2019b}
compared to the Anderson problem on RRG. The first proposal of an
intermediate nonergodic extended phase sandwiched between the deeply
ergodic and insulating (MBL) phases appeared almost 20 years ago \citep{Altshuler1997}.
This phase, colloquially dubbed by Altshuler a ``bad metal'' \citep{Altshuler2010},
was defined as a phase where the eigenfunctions are extended over
the Hilbert space, but cover only $\mathcal{N}^{\gamma}$ states,
where $\gamma<1$ and $\mathcal{N}$ is the Hilbert space dimension.
Whether such an intermediate phase, with multifractal eigenfunctions,
exists for the Anderson localization problem on the Bethe lattice
or RRGs, is still an ongoing debate. Large scale studies on random
regular graphs (RRGs) suggest that this phase disappears in the thermodynamic
limit \citep{Tikhonov2016,Tikhonov2016a,Garcia-Mata2016,Biroli2018,Tikhonov2019},
although there is also no consensus here \citep{Biroli2012,DeLuca2013,Luca,Kravtsov2015,Facoetti2016,Tikhonov2016,Tikhonov2016a,Garcia-Mata2016,Altshuler2016}.
In addition, for weak disorder where all researchers agree that eigenfunctions
are ergodic on RRGs, subdiffusion has been recently observed \citep{PhysRevB.98.134205,de2019sub}.
Notwithstanding, while Anderson localization on graphs and MBL are
related, it is not clear whether results from RRGs apply for MBL.

Multifractal properties of eigenstates of systems which exhibit MBL
where examined in a number of studies \citep{Luitz2015,Torres-Herrera2016,Serbyn2016a,Mace2018,pietracaprina_hilbert_2019}.
The outcome is however rather inconclusive, mostly due to presence
of severe finite size effects (mentioned as well in recent papers
\citep{Panda2019,Abanin2019a,Sierant2019a,Weiner2019a,Khemani2017}).
While Ref.~\citep{Luitz2015} suggests that there is no intermediate
multifractal phase, Refs.~\citep{Pino2015,Torres-Herrera2016} argue
in favor of a stable intermediate phase. In Refs.~\citep{Serbyn2016a,Monthus2016}
multifractal properties of matrix elements of local operators are
studied and found to be multifractal, though Ref.~\citep{Serbyn2016a}
argued that the intermediate phase shrinks to the MBL critical point
in the thermodynamic limit, as it occurs in the standard Anderson
transition. There is therefore a need for a large-scale, numerical
study, which attempts to resolve these discrepancies, and shed light
whether multifractality is related to the anomalous dynamical properties
of the delocalized phase. Two multifractal moments of eigenstates
of the disordered XXZ model were studied in Refs.~\citep{Luitz2015,Mace2018},
and suggest that the extended phase is ergodic. While we see similar
behavior of the relevant moments, in our work we find them insufficient
to unveil possible nonergodic behavior, which becomes only apparent
at higher moments. Our analysis thus allows us to locate a region
in the extended phase which appears to be nonergodic within the available
system sizes.\footnote{This cannot completely rule out strong finite-size effects mentioned
in Refs.~\citep{Panda2019,Abanin2019a,Sierant2019a,Weiner2019a,Khemani2017},
although the presence of multifractal symmetry for the spectrum of
fractal dimensions can be considered as a quite strong argument against
such a scenario.}. Our study also provides, for the first time, the presentation of
the multifractal spectrum. We are able to identify a large portion
of the delocalized phase, where anomalous transport was previously
observed, but which is consistent with a transient multifractality.
Our results support the existence of multifractality in a region which
precedes the MBL transition, although we cannot say whether this region
shrinks to the critical point when the system size is increased (cf.
\citep{Serbyn2016a}).

\section{\label{sec:Model}Model}

In this work we analyze the properties of the disordered XXZ chain,
which is given by the Hamiltonian, 
\begin{equation}
\hat{H}=\frac{J_{xy}}{2}\sum_{i=1}^{L-1}\left(\hat{S}_{i}^{+}\hat{S}_{i+1}^{-}+\hat{S}_{i}^{-}\hat{S}_{i+1}^{+}\right)+J_{z}\sum_{i=1}^{L-1}\hat{S}_{i}^{z}\hat{S}_{i+1}^{z}+\sum_{i=1}^{L}h_{i}\hat{S}_{i}^{z},\label{eq:Model}
\end{equation}
were $\hat{S}_{i}^{z},$ is the $z-$projection of the spin-$1/2$
operator, $\hat{S}_{i}^{\pm}$ are the corresponding lowering and
raising operators, $J_{xy}$ and $J_{z}$ are inter-spin couplings
and $h_{i}$ are random magnetic fields taken to be uniformly distributed
in the interval $h_{i}\in\left[-W,W\right].$ This model conserves
the $z-$projection of the total spin, and serves as the prototypical
model of the MBL transition, which for infinite temperature occurs
for $W\sim W_{c}\backsimeq3.7$ \citep{oganesyan_localization_2007,Berkelbach2010a,Luitz2015}.
For $W\gtrsim W_{c}$ the system is in a MBL phase, with a completely
suppressed transport of all globally conserved quantities \citep{Basko2006a},
while for $W\lesssim W_{c}$ it exhibits an anomalous transport with
a dynamical exponents which depends on the disorder strength \citep{BarLev2014,Lev2014,Agarwal2014,Luitz2015a,Znidaric2016}.
We note in passing that while the value of the critical disorder $W_{c}$
determining the MBL transition in the XXZ Heisenberg model is under
debate (cf. $W\backsimeq3.7$ in Refs. \citep{oganesyan_localization_2007,Berkelbach2010a,Luitz2015}
vs $W\gtrsim4.5$ in Refs.~\citep{Devakul2015,Khemani2017,Doggen2018,Weiner2019a}),
since one of the objectives of this work is to study the connection
between anomalous transport and multifractality to avoid the controversy
we limit the disorder strengths in our study to , $W\leq3$, which
according to all studies belong to the delocalized phase.

\section{Results}

Multifractal analysis requires the calculation of the eigenstates
of (\ref{eq:Model}) in a certain energy density window and for a
large number of disorder realizations. Since full diagonalization
becomes overwhelmingly expensive for system sizes $L\gtrsim18$, and
access to large system sizes is essential, we utilize the shift-invert
technique \citep{Pietracaprina2018}, which transforms the spectrum
of the Hamiltonian such that the states of interest are moved to the
lowest (highest) energies in the transformed spectrum and become tractable
by Krylov space methods. The most commonly used spectral transformation
for this purpose is $\left(H-\sigma I\right)^{-1}$, where the explicit
inversion of the shifted Hamiltonian can be avoided and replaced by
the solution of a set of linear equations using the Gauss algorithm.
We use the massively parallel \texttt{strumpack} library \citep{ghysels_efficient_2016,ghysels_robust_2017}
to extract about 50 eigenstates in the middle of the many-body spectrum,
where the density of states is at its maximum. The largest system
size we consider is $L=24$, which corresponds to a Hilbert space
dimension of ${\cal N}=2\,704\,156$. We repeat this procedure for
$100-15\,000$ realizations of the disordered magnetic field $h_{i}$
in (\ref{eq:Model}). Overall, the total number of calculated eigenstate
coefficients in the computational basis for each disorder strength
and system size is $10^{8}-10^{10}$, which in most cases allows us
to reach statistical errors smaller than the symbol size.

\subsection{\label{subsec:Distributions-of-eigenstate}Distributions of eigenstate
coefficients}

The high energy states of \emph{ergodic} systems are well approximated
by eigenstates of random-matrices drawn from a Wigner-Dyson ensemble
of matrices \citep{mehta} which shares the same temporal symmetry
as the Hamiltonian. Specifically, eigenstates of real ergodic Hamiltonians,
which are time-inversion invariant, are well described by eigenstates
of matrices drawn from the Gaussian Orthogonal Ensenble (GOE), suggesting
that the elements of eigenstates, $\ket{\beta}$, written in a certain
basis $\ket n$, are almost independent random variables, normally
distributed according to, 
\begin{equation}
P_{\text{GOE}}\left(x\equiv\left|\left\langle n|\beta\right\rangle \right|\right)=\sqrt{\frac{\mathcal{N}}{2\pi}}e^{-\mathcal{N}x^{2}/2},\label{eq:Berry_conjecture}
\end{equation}
where we defined the random variable, $x\equiv\left|\left\langle n|\beta\right\rangle \right|$
and $\mathcal{N}$ is the Hilbert space dimension. This assertion,
known as Berry's conjecture \citep{Berry1977}, was verified numerically
in several single- and many-body ergodic systems (see Ref.~\citep{Guhr1998}
for a review). Multifractal eigenstates, on the contrary, do \emph{not}
satisfy Berry's conjecture, but are distributed according to, 
\begin{equation}
P_{\N}\left(x\right)\propto\frac{1}{\left|x\right|}\N^{f\left(-\ln x^{2}/\ln\N\right)-1},\label{eq:f_alpha_definion}
\end{equation}
where $f\left(\alpha\right)$ is a function called the spectrum of
fractal dimensions~\citep{Evers2008a}, depending on the only variable
$\alpha$ taken to be $\alpha\equiv-\ln x^{2}/\ln\N$ (see Section~\ref{subsec:Multrifractal-spectrum},
and Eq.~(\ref{eq:f_alpha_GOE_N}) for the form of $f\left(\alpha\right)$
for GOE eigenstates).

The distribution of eigenstate coefficients for our model (\ref{eq:Model})
has been studied by two of us in Ref.~\citep{Luitz2016b}, and was
found to exhibit significant deviations from Berry's conjecture for
$0.4\leq W\leq1.8$, hinting that the underlying eigenstates are multifractal.
In this work we study these distributions in detail, focusing on their
flow towards the thermodynamic limit. In both above cases we focus
on eigenstate coefficients in the computational basis, where the basis
states $\ket n$ are labeled by the eigenvalues of the local $\hat{S}_{i}^{z}$
operators.

\begin{figure}
\includegraphics[width=0.9\textwidth]{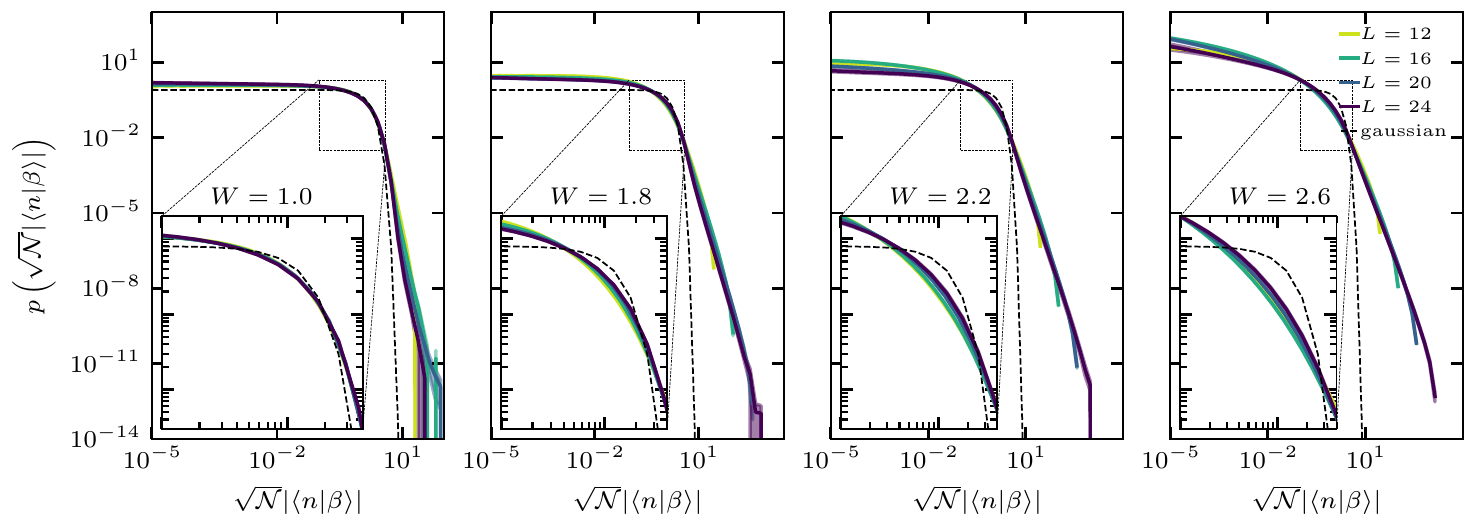}

\caption{\label{fig:P_abs_psi_rescaled}Normalized probability density of scaled
eigenstate coefficients in the computational basis, $P\left(\sqrt{\mathcal{N}}\left|\left\langle n|\beta\right\rangle \right|\right)$
for disorder strengths $W=1.0,1.8,2.2$ and $2.6$ and system sizes
$L=12,16,20$ and 24 (larger systems correspond to darker colors).
The dashed black line represents the normal distribution and errorbars
are represented by shaded areas of the order of the line width.}
\end{figure}

To give equal weight to small and large values of the eigenstate coefficients,
the bins of the histogram are equally spaced on a \emph{logarithmic}
scale, which we achieve by calculating the histogram of $\alpha\equiv-\ln\left|\left\langle n|\beta\right\rangle \right|^{2}/\ln\N$
using bins of equal size. The histogram of the wavefunction coefficients
$|\langle n|\beta\rangle|$ and the corresponding probability density
can then be straightforwardly inferred. In Fig.~\ref{fig:P_abs_psi_rescaled}
we show the result of this calculation for a number of disorder strengths
in the extended phase, and a range of system sizes. We compare these
distributions with the normal distribution of GOE, (\ref{eq:Berry_conjecture}),
and see a visible departure for all disorder strengths, similarly
to Fig.~2 in Ref.~\citep{Luitz2016b}. The departure is especially
apparent in the head of the distribution, indicating an excess in
small values of the eigenstate elements compared to GOE, and the tails
of the distribution, indicating an excess in large values of the eigenstate
elements. The departure becomes more prominent with the strength of
the disorder.

While at first glance the rescaled distributions look collapsed, a
more detailed examination by zooming into various parts of the distribution,
shows a noticeable, yet slow, flow towards the (Gaussian) GOE distribution
in most parts of the distribution. In what follows we examine this
flow in detail, by considering the moments of the distribution and
its multifractal spectrum (\ref{eq:f_alpha_definion}).

\subsection{Moments of the distributions: inverse participation ratios}

\begin{figure}[th]
\includegraphics[width=0.9\textwidth]{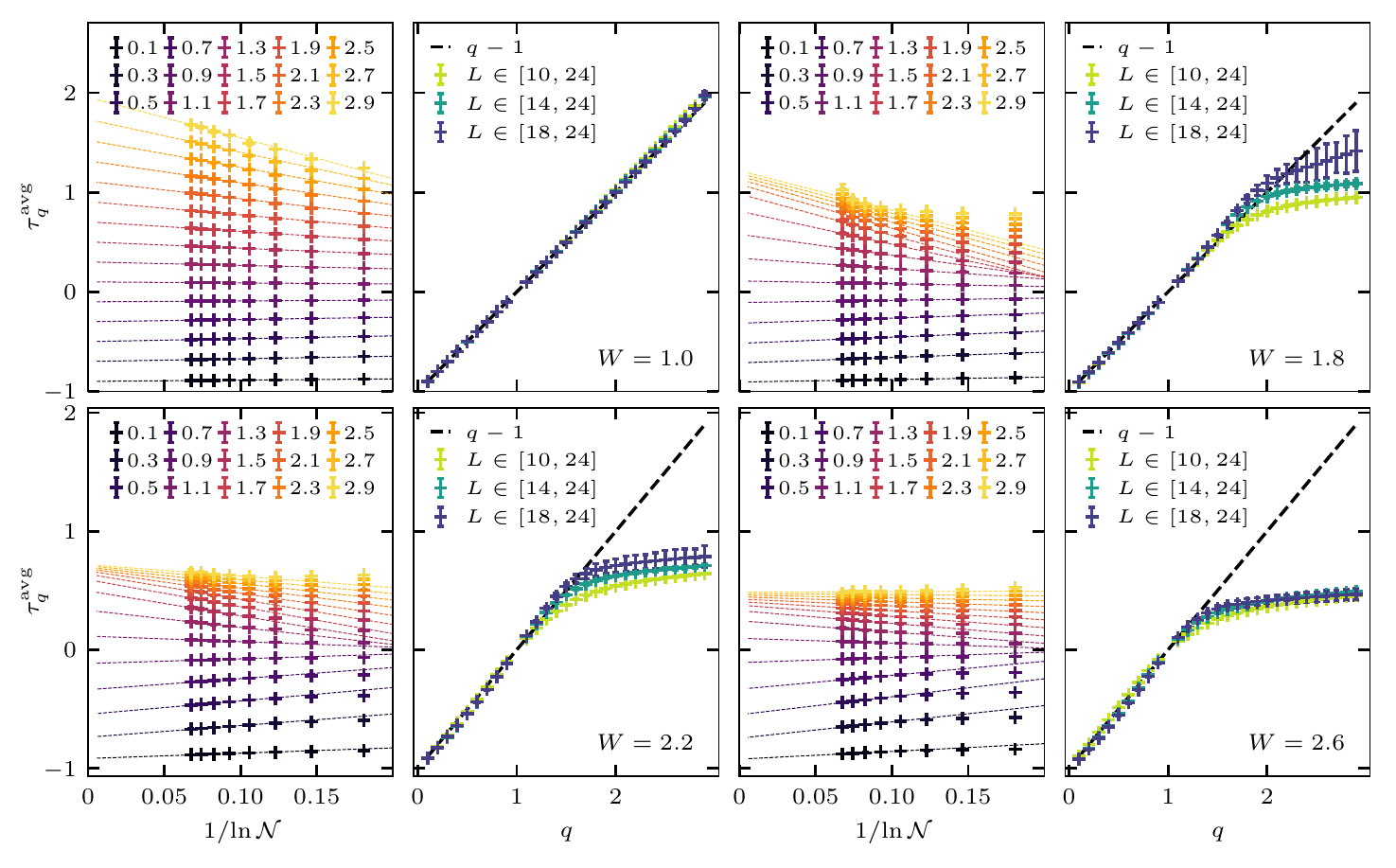}

\caption{\emph{Odd columns}. Finite size $\tau_{q}^{\protect\N,\text{avg}}$
data (\ref{eq:tau_q_avg}) as a function of $1/\ln\protect\N$ for
various values of $q$ ($0<q<3$, darker shade of green indicates
smaller $q$) and disorder strengths $W=1.0,1.8,2.2$ and $2.6$.
Statistical errors are smaller than symbol size in all plots. Dashed
lines are extrapolation of the data to $\protect\N\to\infty$ using
a linear function in $1/\ln\mathcal{N}$. \emph{Even columns}. Extrapolated
$\tau_{q}^{\text{avg}}$ as a function of $q$ and various extrapolation
ranges indicated in the legend. Darker colors indicate more weight
to larger system sizes. The dashed black line indicates $q-1$, which
corresponds $\tau_{q}^{\text{avg}}$ of the GOE ensemble.}
\label{fig:tau_q_avg}
\end{figure}

In the multifractal analysis one defines the standard inverse participation
ratio (IPR) $I_{2}^{\beta}$ and its generalizations, 
\begin{equation}
I_{q}^{\beta}=\sum_{n}\left|\left\langle \beta|n\right\rangle \right|^{2q}\sim\mathcal{N}^{-\tau_{q}},\label{eq:i_q}
\end{equation}
which measure how many ``sites'' in the Hilbert space (a site here
is a certain basis state $\ket n$) the wavefunction occupies \citep{Evers2008a},
the generalized IPR is directly related to the corresponding $q$
Rényi entropies $S_{q}=\ln I_{q}^{\beta}/(1-q)$ \citep{Luitz2014}.
For eigenstates extended over the entire basis, such as for eigenstates
extracted from GOE, $x\equiv\left|\left\langle \beta|n\right\rangle \right|\sim\mathcal{N}^{-1/2}$
giving, $I_{q}^{\beta}=\sum_{n}\mathcal{N}^{-q}=\mathcal{N}^{-\left(q-1\right)}$
and $\tau_{q}=q-1$ for $q>-0.5$ (the average of $I_{q}^{\beta}$
diverges otherwise). Eigenstates which occupy a finite number of configurations
$\ket n$ which doesn't scale with the Hilbert space dimension $\mathcal{N}$
will have $\tau_{q}=0$ for $q>0$ (and $-\infty$ otherwise). The
parameter $q$ is used to tune the weight in the average from large
to small values. Under the assumption that the eigenstate coefficients
are statistically independent, the IPRs are related to the moments
of $P\left(x\right)$, since one can write, 
\begin{equation}
I_{q}^{\beta}=\sum_{n}\left|\left\langle \beta|n\right\rangle \right|^{2q}=\mathcal{N}\frac{1}{\mathcal{N}}\sum_{n}\left|\left\langle \beta|n\right\rangle \right|^{2q}\approx\mathcal{N}\left\langle \left|x\right|^{2q}\right\rangle =\mathcal{N}\int\left|x\right|^{2q}P\left(x\right)\mathrm{d}x.\label{eq:i_q_rel_to_moments}
\end{equation}
Using the definition of $\tau_{q}$ in (\ref{eq:i_q}), the normalization
of the wavefunction, which gives, $I_{q=1}^{\beta}=1$, and the fact
that $\sum_{n}1=\N$, which gives $I_{q=0}^{\beta}=\N$ one can show
that in the limit of $\mathcal{N}\to\infty$, $\tau_{q}$ is a monotonically
increasing and concave function of $q$, namely $\tau_{q}'>0$ and
$\tau_{q}''<0$ \citep{Evers2008a}.

To evaluate the $\tau_{q}$ we calculate the IPRs for each eigenfunction
and a range $0\leq q\leq4$. We then average $I_{q}^{\beta}$ over
the nearby in energy eigenstates, as also different disorder realizations,
and obtain $\left\langle I_{q}^{\mathcal{N}}\right\rangle $. The
finite-size \emph{average} $\tau_{q}^{\text{avg}}$ is the given by,
\begin{equation}
\tau_{q}^{\text{avg}}\left(\N\right)\equiv-\frac{\ln\left\langle I_{q}^{\mathcal{N}}\right\rangle }{\ln\N}.\label{eq:tau_q_avg}
\end{equation}
\begin{figure}
\includegraphics[width=0.9\textwidth]{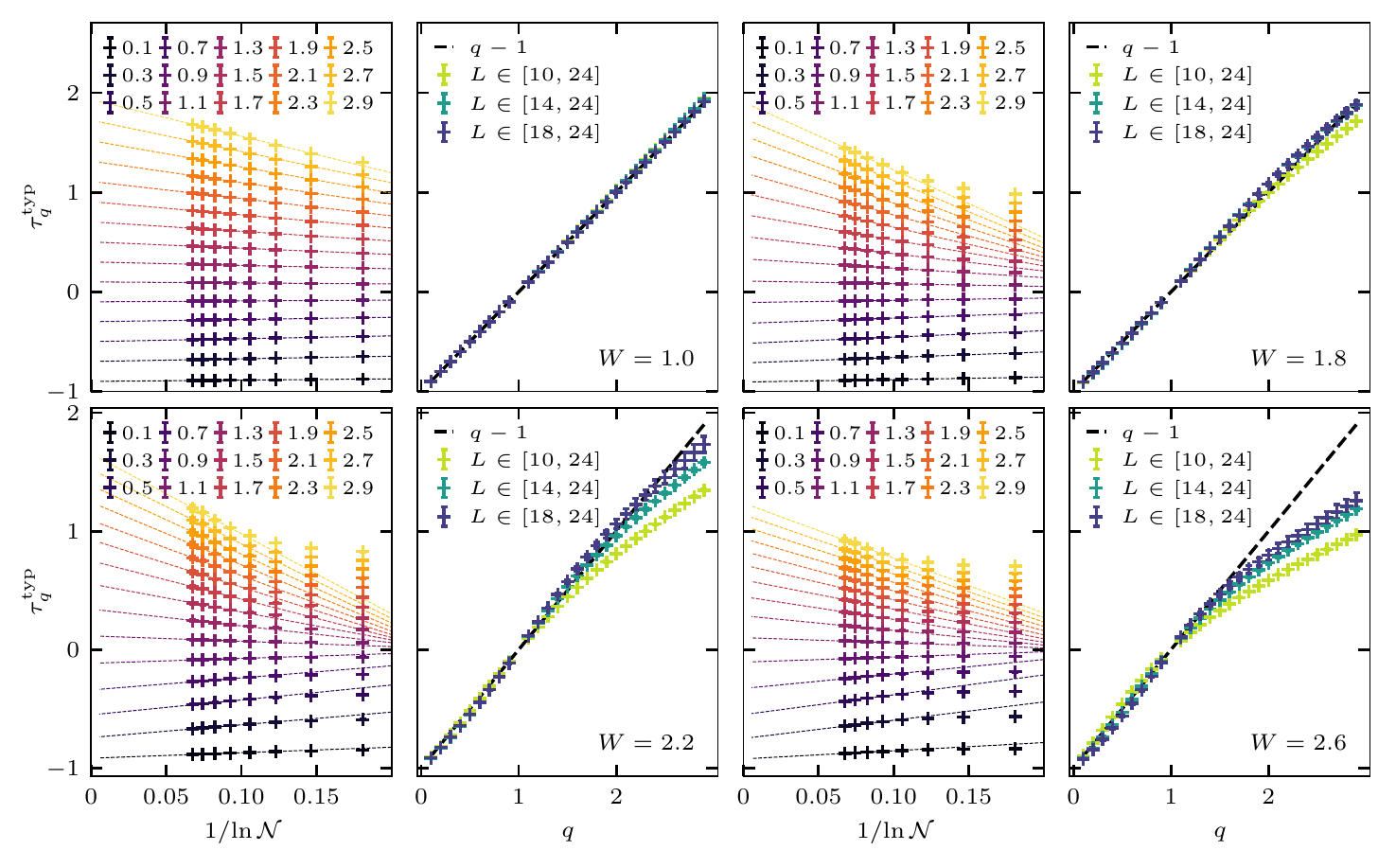}

\caption{Similar to Fig.~\ref{fig:tau_q_avg}, but for $\tau_{q}^{\text{typ}}$,
computed from Eq.~(\ref{eq:tau_q_typ}).}
\label{fig:tau_q_typ}
\end{figure}
Since the relation (\ref{eq:i_q}) is only expected to hold asymptotically,
in Fig.~\ref{fig:tau_q_avg} we plot $\tau_{q}^{\text{avg}}\left(\N\right)$
vs $1/\ln\N$ and extrapolate to $\N\to\infty$ using a linear function
in $1/\ln{\cal N}$ \footnote{The motivation of the extrapolation versus {$1/\ln\N$} originates
from the main subleading contributions to {$\tau_{q}\left(\N\right)$}
from the prefactor in (\ref{eq:i_q}).}. The extrapolated values are then plotted as a function of $q$ and
compared to the prediction of GOE, $\tau_{q}=q-1$ (dashed black line).
While the scaling of $\tau_{q}^{\text{avg}}\left(\N\right)$ with
respect to $1/\ln\N$ is mostly linear indicating a high quality of
the extrapolation, for larger values of $q$ a departure from the
linear dependence is apparent, suggesting that the data is still far
from being asymptotic. The slight non-concavity of the extrapolated
$\tau_{q}^{\text{avg}}$, is a finite size effect and is well within
the error bars of the extrapolation. To quantify the curving of the
data, we extrapolate to $\N\to\infty$ using a sliding fit window
of system sizes, which are shown in the legend. The error bars in
the extrapolated data are estimated using a bootstrap fitting procedure,
quantifying the statistical errors in $\tau_{q}^{\text{avg}}\left(\N\right)$
(which are in all cases smaller than the symbol size). From the extrapolated
data we see that the average, $\tau_{q}^{\text{avg}}\sim q-1$ for
all values of $q$ up to some $q_{*}\left(W,L\right)$, which depends
on the strength of the disorder. While for weak disorder $q_{*}$
spans the entire range of $q$ considered here, for $W\to W_{c}$
we see that $q_{*}\to1$ . On the other hand we also see that, $q_{*}\left(W,L\right)$
increases when higher weight in the extrapolation is given to the
larger system sizes, suggesting that the departure from the $q-1$
could be a finite size effect, though we cannot rule out a saturation
of the form $\lim_{L\to\infty}q_{*}\left(W,L\right)=q_{*}\left(W\right)$,
which will indicate residual multifractality at large moments $q>q_{*}\left(W\right)$.

We also study the typical $\tau_{q}^{\text{typ}}\left(\N\right)$,
which is defined as 
\begin{equation}
\tau_{q}^{\text{typ}}\left(\N\right)\equiv-\frac{\ln\left\langle I_{q}^{\mathcal{N}}\right\rangle _{\text{typ}}}{\ln\N},\qquad\text{where }\left\langle I_{q}^{\mathcal{N}}\right\rangle _{\text{typ}}\equiv\exp\left[\left\langle \ln I_{q}\right\rangle \right].\label{eq:tau_q_typ}
\end{equation}
The advantage of this measure is that it suppresses the weight of
the outliers. The results of the same analysis for $\tau_{q}^{\text{typ}}$
as described above for $\tau_{q}^{\text{avg}}$ are presented in Fig.~\ref{fig:tau_q_typ},
and are qualitatively identical to the analysis of $\tau_{q}^{\text{avg}}$.
Quantitatively $q_{*}$ is pushed to much larger values~\footnote{as in the GOE case at finite size, where {$q^{*}\sim\ln\N$}, see~\citep{Baecker2019}
for details.}, almost entirely eliminating the departure from the $q-1$ line for
all $W<2.6$. This can be viewed as another indication that $q_{*}$
is dominated by outliers and is likely to flow to infinity for larger
system sizes.

\begin{figure}
\includegraphics[width=0.9\textwidth]{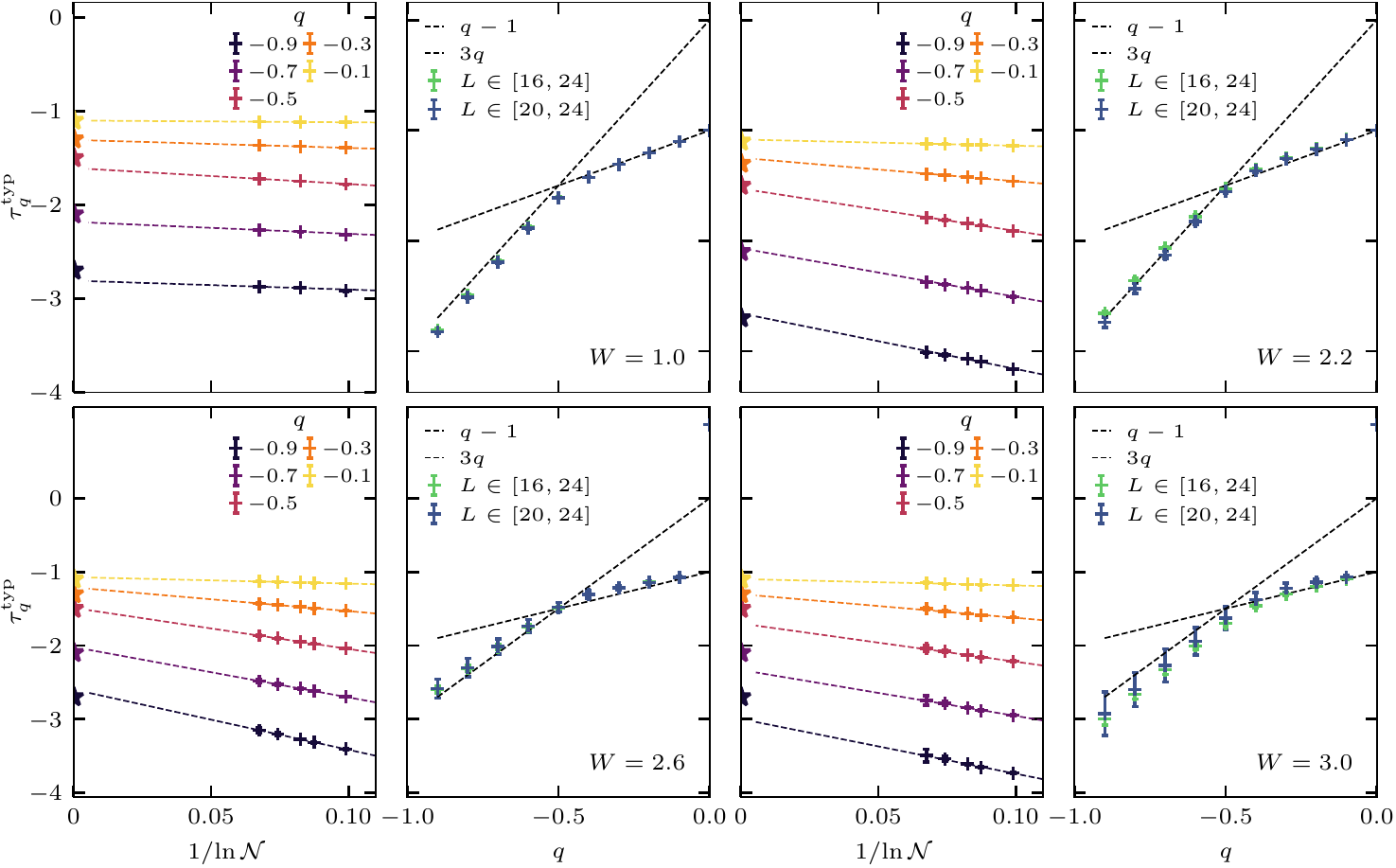}

\caption{\label{fig:fig_tau_typ_neg}Similar to Fig.~\ref{fig:tau_q_typ}
but for $W=1.0,2.2,2.6$ and 3.0 for $q<0$. The dashed black lines
correspond to the expected behavior of $\tau_{q}^{\text{typ}}$ for
the GOE case, $3q$ for $q<-0.5$ and $(q-1)$ otherwise. See the
Eq.~(\ref{eq:tau_q_typ_GOE}) for details.}
\end{figure}

Another advantage of $\tau_{q}^{\text{typ}}$ is that unlike $\tau_{q}^{\text{avg}}$
it doesn't diverge for $q<-0.5$, and thus allows to study the behavior
of the small values of the eigenstate coefficients. We recall that
these values are of particular interest given their abundance compared
to the Gaussian distribution (see Fig.~\ref{fig:P_abs_psi_rescaled})\textcolor{red}{{}
}. In Fig.~\ref{fig:fig_tau_typ_neg} we repeat the analysis done
in Fig.~\ref{fig:tau_q_typ} for $q<0$ (for technical reasons we
use a different set of data here, which includes less samples). 

We estimate the value of $\tau_{q}^{\text{typ}}$ for GOE eigenstates
based on the behavior of $f_{\text{avg}}^{\text{GOE}}\left(\alpha\right)$
for $\alpha>1$, which can be calculated analytically based on (\ref{eq:Berry_conjecture})
and (\ref{eq:f_alpha_definion}), 
\begin{equation}
f_{\text{av}}^{\text{GOE}}\left(\alpha\right)=\begin{cases}
-\infty & \alpha<1\\
\left(3-\alpha\right)/2 & \alpha>1
\end{cases}.\label{eq:f_alpha_GOE}
\end{equation}
Since the typical $f_{\text{typ}}\left(\alpha\right)$ is determined
by histogram counts growing with the system size~\citep{Evers2008a},
it coincides with $f_{\text{avg}}\left(\alpha\right)$ for $f_{\text{avg}}\left(\alpha\right)\geq0$
and tends to $-\infty$ (zero counts) otherwise. Thus, we can evaluate
$\tau_{q}^{\text{typ}}$ from $f_{\text{av}}^{\text{GOE}}\left(\alpha\right)$
using a truncated Legendre transform \citep{Evers2008a}, 
\begin{equation}
\tau_{q}^{\text{typ}}=\sup_{{\alpha\atop f_{\text{av}}^{\text{GOE}}\left(\alpha\right)>0}}\left[q\alpha-f_{\text{av}}^{\text{GOE}}\left(\alpha\right)\right]=\begin{cases}
q-1 & q>-0.5\\
3q & q\leq-0.5
\end{cases},\label{eq:tau_q_typ_GOE}
\end{equation}
which is designated in Fig.~\ref{fig:fig_tau_typ_neg} by the dashed
black lines. We note that similarly to $q>0$, $\tau_{q}^{\text{typ}}$
for $q<0$ appears to flow to the predictions of GOE. 
\begin{figure}
\includegraphics[width=0.9\textwidth]{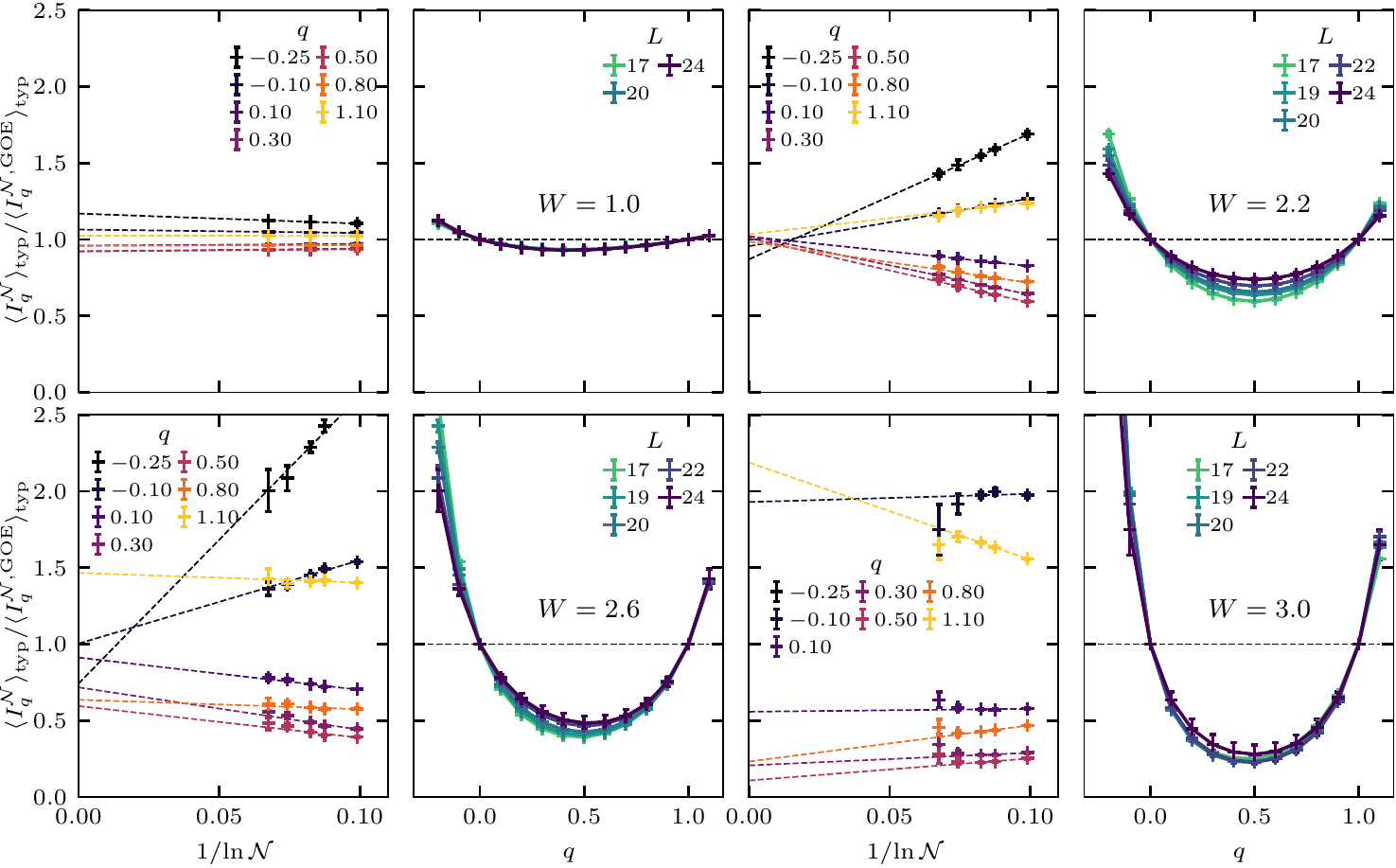}

\caption{\label{fig:iq_vs_iq_goe}\emph{Odd columns.} Finite size data of the
ratio, $\left\langle I_{q}^{\text{ }\mathcal{N}}\right\rangle _{\text{typ}}/\left\langle I_{q}^{\protect\N,\text{GOE}}\right\rangle _{\text{typ}}$
vs $1/\ln\protect\N$ for a number of $q$-s and disorder strength
$W$ (see legend). Dashed lines are extrapolation to $\protect\N\to\infty$.
\emph{Even columns}. The same ratio $\left\langle I_{q}^{\text{ }\mathcal{N}}\right\rangle _{\text{typ}}/\left\langle I_{q}^{\protect\N,\text{GOE}}\right\rangle _{\text{typ}}$,
but as a function of $q$. Dashed black line indicates GOE limit.}
\end{figure}

This conclusion is in apparent contradiction to the behavior of the
distributions in Fig.~\ref{fig:P_abs_psi_rescaled}, where an excess
of zeros of the eigenstates compared to GOE prediction is clearly
visible for $W\sim1$, and does not appear to vanish in the $\N\to\infty$.
The discrepancy must follow from the prefactor in the definition of
$\tau_{q}^{\text{typ}}$, $\left\langle I_{q}^{\text{ }\mathcal{N}}\right\rangle _{\text{typ}}=A\left(\N\right)\N^{-\tau_{q}^{\text{typ}}}$,
which includes a slowly varying prefactor (the variation is at most
of the order of $\ln\N$). To test this assertion numerically, we
compute the ratio, $\left\langle I_{q}^{\text{ }\mathcal{N}}\right\rangle _{\text{typ}}/\left\langle I_{q}^{\N,\text{GOE}}\right\rangle _{\text{typ}}$
for a number of $q$-s and disorder strengths. Since to the best of
our knowledge there are no analytical relations for $\left\langle I_{q}^{\N,\text{GOE}}\right\rangle _{\text{typ}}$~\footnote{Unlike for the average {$\left\langle I_{q}^{\N,\text{GOE}}\right\rangle $},
see, e.g., \citep{Brody1981}.} we compute it numerically by drawing 100 random eigenstates from
a Gaussian probability distribution in Eq.~(\ref{eq:Berry_conjecture}),
while fixing the norm of the eigenstate. This procedure is very efficient
and allows us to study the same Hilbert space dimensions as we do
for the XXZ model, at a negligible computational cost. The results
of the evaluation of $\left\langle I_{q}^{\text{ }\mathcal{N}}\right\rangle _{\text{typ}}/\left\langle I_{q}^{\N,\text{GOE}}\right\rangle _{\text{typ}}$
can be seen in Fig.~\ref{fig:iq_vs_iq_goe}. For $W<1$ we indeed
see that for the system sizes we have the ratio flows away from 1.
We strongly suspect that this apparent divergence from GOE, is a finite
size effect, which has to do with the proximity of an integrable point
(for $W=0$ the XXZ model is integrable). We leave the examination
of this effect to future studies. For $W>1$ we see an apparent flow
towards 1, with clearest evidence for $W=2.2$. For $W=2.6$ and 3.0,
the finite size behavior is non-monotonic (highlighting the importance
of the use of large system sizes), and appears to flow towards 1,
though for these disorder strengths it is less apparent.

To summarize this section, we have seen that while a naïve examination
of the distributions of the eigenstates elements in Fig.~\ref{fig:P_abs_psi_rescaled},
shows apparent convergence to a non-Gaussian, and thus multifractal
distribution, a more detailed analysis looking on the moments of the
distribution, shows a slow but clear flow towards the predictions
of GOE, in $\tau_{q}^{\text{typ}},\tau_{q}^{\text{avg}}$ and even
directly in the finite size generalized IPRs compared to thier GOE
values $\left\langle I_{q}^{\text{ }\mathcal{N}}\right\rangle _{\text{typ}}/\left\langle I_{q}^{\N,\text{GOE}}\right\rangle _{\text{typ}}$.
In the next section we will complement this analysis, by examining
an additional multifractal measure --- the multifractal spectrum.

\subsection{\label{subsec:Multrifractal-spectrum}Multifractal spectrum}

\begin{figure}
\includegraphics[width=0.9\textwidth]{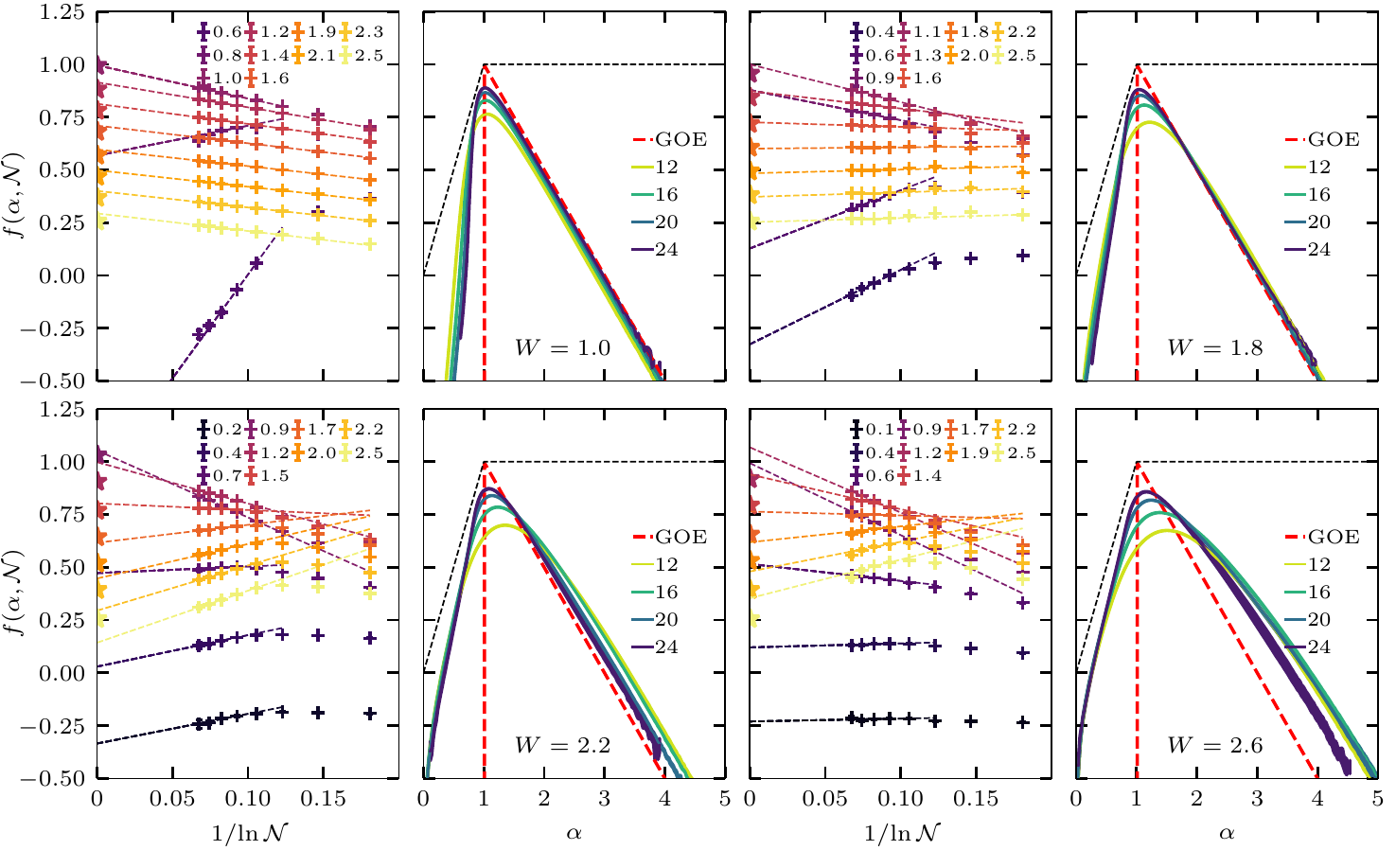}

\caption{\emph{Odd columns}. $f\left(\alpha,\protect\N\right)$ vs $1/\ln\protect\N$
for different $\alpha$ values (see legend). Dashed colored lines
correspond to an extrapolation to $\protect\N\to\infty$, and the
stars on the y-axis correspond to the infinite size GOE prediction.
\emph{Even columns}. The finite size multifractal spectrum, $f\left(\alpha,\protect\N\right)$
calculated using Eq.~(\ref{eq:f_alpha_N}) for $L=12,16,20,24$ (darker
colors, correspond to larger system sizes) and disorder strengths
$W=1.0,1.8,2.2,2.6$, the width of the lines correspond to statistical
errors. Red dashed line corresponds to the infinite size GOE prediction,
$f_{\text{GOE}}\left(\alpha\right)$ according to (\ref{eq:f_alpha_GOE_N}),
and black dashed lines show the upper bounds on $f\left(\alpha\right)$
according to (\ref{eq:f_alpha_bounds}). The error bars are represented
by filled areas of the order of the line width.}
\label{fig:falpha}
\end{figure}

In this section we analyze the multifractal spectrum, $f\left(\alpha\right)$,
of the eigenstates of (\ref{eq:Model}), which appeared in (\ref{eq:f_alpha_definion}),
but we repeat it here for convenience, 
\begin{equation}
P_{\N}\left(x\right)\propto\frac{1}{\left|x\right|}\N^{f\left(-\ln x^{2}/\ln\N\right)-1},\label{eq:f_alpha_definion-1}
\end{equation}
with $\alpha\equiv-\ln x^{2}/\ln\N$ and $x\equiv\left|\left\langle n|\beta\right\rangle \right|$,
where $\ket n$ is a basis state, and $\ket{\beta}$ are eigenstates
of (\ref{eq:Model}) \citep{Evers2008a}. The multifractal spectrum,
$f\left(\alpha\right)$ is the fractal dimension of the set $x\in\left[\N^{-\alpha/2},\N^{-\left(\alpha+\mathrm{d}\alpha\right)/2}\right]$,
namely the probability for $x$ to be in this interval is given by,
\begin{equation}
p\left(\alpha\right)\sim\left(\ln\N\right)\N^{-\left(1-f\left(\alpha\right)\right)}.\label{eq:prob_psi_log_bins}
\end{equation}
Using (\ref{eq:f_alpha_definion-1}) and the relation (\ref{eq:i_q_rel_to_moments})
to $I_{q}$ one can see that, 
\begin{equation}
\tau_{q}=\inf_{\alpha}\left[q\alpha-f\left(\alpha\right)\right],
\end{equation}
namely in the limit $\N\to\infty$, $\tau_{q}$ and $f\left(\alpha\right)$
are related via a Legendre transform \citep{Evers2008a}. Here $\inf$
is an infimum. We note however that while $\tau_{q}$ is a concave
function the definition (\ref{eq:f_alpha_definion-1}) above allows
for $f\left(\alpha\right)$ to be \emph{non-concave} a feature which
we will utilize in our analysis below. Similarly to the restrictions
on $\tau_{q}$, described above Eq.~(\ref{eq:i_q_rel_to_moments}),
from normalization of the probability distribution, $\int P_{\N}\left(x\right)\mathrm{d}x=1$
and the wavefunction $\int x^{2}P_{\N}\left(x\right)\mathrm{d}x=\N^{-1}$
in the limit $\N\to\infty$ one can derive, that 
\begin{equation}
f\left(\alpha\right)\le\min\left(1,\alpha\right).\label{eq:f_alpha_bounds}
\end{equation}
For the Gaussian distribution of GOE eigenstates (\ref{eq:Berry_conjecture}),
using the definition (\ref{eq:f_alpha_GOE}), one can obtain the finite
size correction, 
\begin{equation}
f_{\text{GOE}}\left(\alpha,\N\right)=1+\frac{\ln\left(P\left(\alpha\right)/\left(A\ln\N\right)\right)}{\ln\N}=1+\frac{1-\alpha}{2}-\frac{\N^{1-\alpha}}{2\ln\N}-\frac{\ln A}{\ln\N},\label{eq:f_alpha_GOE_N}
\end{equation}
which in the limit $\N\to\infty$ gives the already mentioned result
(\ref{eq:f_alpha_GOE}). Here $A$ is a normalization constant being
a slow (at most logarithmic) function of $\N$. Note that this multifractal
spectrum differs from $f_{\text{GOE}}\left(1\right)=1$, $f_{\text{GOE}}\left(\alpha\ne1\right)\to-\infty$
in Ref.~\citep{Evers2008a} because the latter is written for wavefunction
envelopes, while the raw numerical eigenstates contain de~Broglie-like
oscillations corresponding to the increased statistics of zeros (large
values of $\alpha>1$). While there are methods to remove these superfluous
zeros (see, e.g.,~\citep{Luca,Kravtsov2015}), since the statistics
of the zeros does not affect $q>0$ moments of the eigenfunctions
we don't consider such methods in this work.

To obtain the finite size multifractal spectrum we numerically compute
a histogram of $\alpha$ with $0\leq\alpha\leq4$ (we used 50, 100,
and 200 bins, and verified that our results don't change with respect
to the bin number, not shown), then using (\ref{eq:prob_psi_log_bins})
yields, 
\begin{equation}
f\left(\alpha,\N\right)=1+\frac{\ln\left(p\left(\alpha\right)/\ln\N\right)}{\ln\N},\label{eq:f_alpha_N}
\end{equation}
which is presented in the even columns of Fig.~\ref{fig:falpha},
for various disorder strengths $1\leq W\leq2.6$. Here we assumed
$A$ to be a constant (not a logarithmic function of $\N$) like in
the GOE case as we focus on the small disorder amplitudes. In the
odd columns of Fig.~\ref{fig:falpha} we extrapolate the data to
$\N\to\infty$, with the same procedure used in several random matrix
models (see, e.g., \citep{Luca,Kravtsov2015}). For sufficiently strong
disorder or $\alpha$ sufficiently far from 1, our finite size data
shows a nonlinear behavior in $1/\ln\N$ (like in GOE case (\ref{eq:f_alpha_GOE}))
indicating the importance of using large system sizes in the the determination
of the asymptotic behavior. Even with the state-of-the-art system
sizes we use here, the extrapolation procedure is not justified due
to nonlinearity of the data for some values of $\alpha$. Nevertheless,
for $W<2$, the extrapolation works fairly well, and similarly to
the moments analysis in the previous section, supports a flow towards
the predictions of GOE. The extrapolation is not entirely satisfactory
for $W=2.2$ and 2.6, thus we cannot rule out multifractal behavior
in this case.

We also note that in the calculation of the histograms of $\alpha$,
the sampling for our system sizes starts yielding zero counts (for
all used bin sizes) for $\alpha\gtrsim2.5$ and the majority of eigenvectors.
In this interesting regime (corresponding to the excess of zeros in
the wavefunction histogram, Fig.~\ref{fig:P_abs_psi_rescaled}),
the (low probability) contribution to the histogram seems to stem
from the distribution over disorder realizations, rather than from
representative eigenstates. This leads to large fluctuations, also
visible in the errorbars (shaded area).

\begin{figure}[th]
\includegraphics[width=0.9\textwidth]{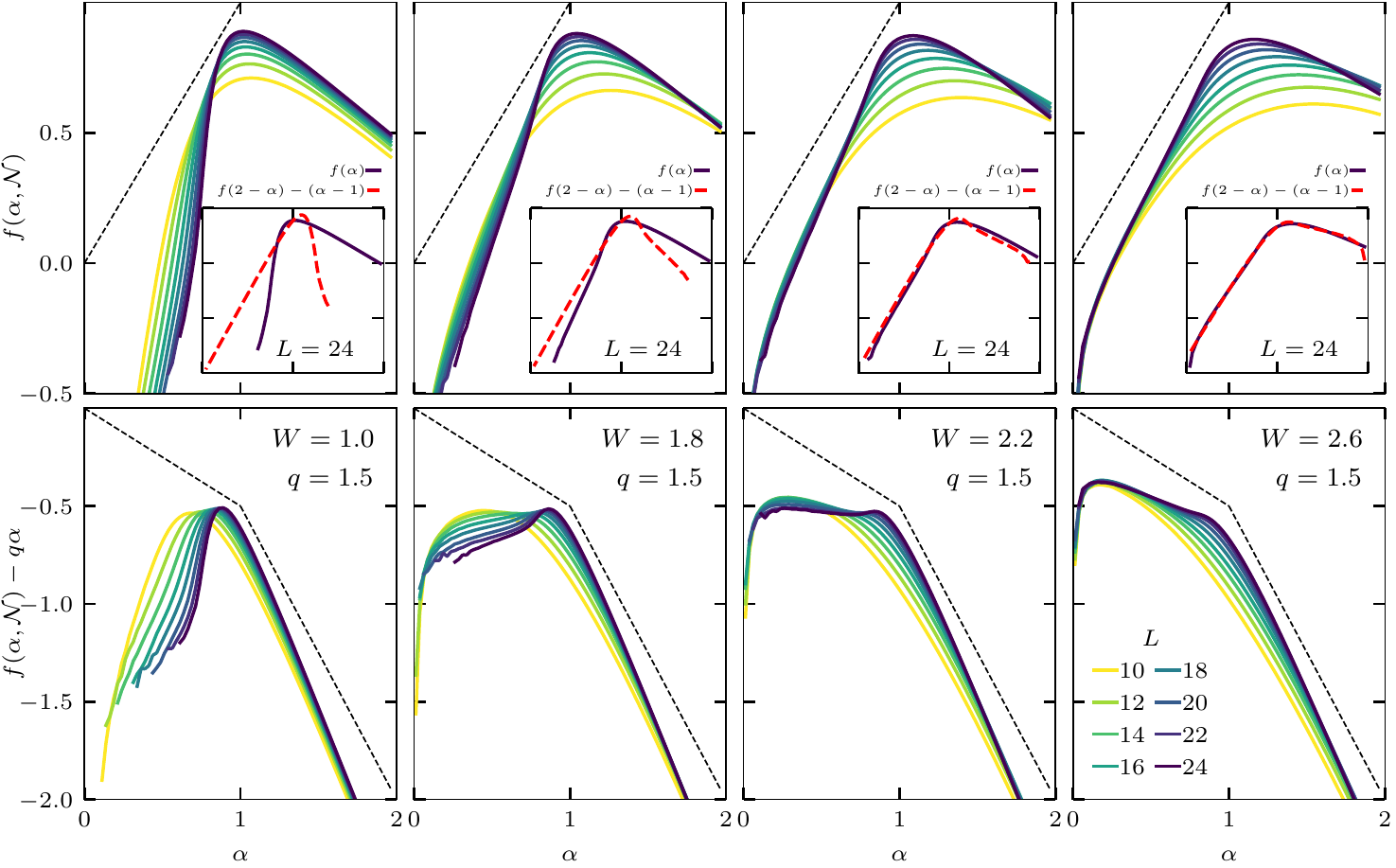}

\caption{\emph{Upper row}. The finite size multifractal spectrum $f\left(\alpha,\protect\N\right)$
for small values of $\alpha$ and $L=12,16,20,24$ (darker colors,
correspond to larger system sizes). Dashed black lines indicate the
upper bounds on $f\left(\alpha\right)$ according to (\ref{eq:f_alpha_bounds}).
The insets show $f\left(\alpha,\protect\N\right)$ (solid black lines)
for the maximal available size $\left(L=24\right)$ together with
its multifractally symmetric counterpart according to Eq. (\ref{eq:MF-symmetry})
(dashed red lines). \emph{Lower row}. Same as the upper row, but for
the tilted multifractal spectrum, $f\left(\alpha,\protect\N\right)-q\alpha$,
and $q=1.5$.}
\label{fig:falpha-qalpha}
\end{figure}

The finite size behavior of the multifractal spectrum, $f\left(\alpha,\N\right),$
is also useful to understand the deviation from the $\left(q-1\right)$
GOE line for both $\tau_{q}^{\text{avg}}$ and $\tau_{q}^{\text{typ}}$,
which occurs for some value $q_{*}\left(W,L\right)$ (see Figs.~\ref{fig:tau_q_avg}
and \ref{fig:tau_q_typ}). By looking on the direction of the flow
of $f\left(\alpha,\N\right)$ with the system size (see Fig.~\ref{fig:falpha}),
for all $W\leq2.2$ and $\alpha$ close to the maximum of $f\left(\alpha,\N\right)$,
which corresponds to $q<q^{*}$, the $f\left(\alpha,\N\right)$ \emph{increases}
with $\N$. For small $\alpha$ on the other hand, which corresponds
to $q>q^{*}$, the spectrum $f\left(\alpha,\N\right)$ \emph{decreases}
with $\N$. Moreover the crossing points of $f\left(\alpha,\N\right)$
at two adjacent $\N$ values flow towards $\alpha=1$ with increasing
$\N$. For $W=2.6$ the situation is drastically different, since
there is no downward finite-size flow at small $\alpha$, but instead
$f\left(\alpha,\N\right)$ appears to saturate. This is also visible
in the extrapolation curves for small $\alpha=0.1$ in Fig.~\ref{fig:falpha}.
To emphasize this point in Fig.~\ref{fig:falpha-qalpha} we plot
$f(\alpha)-q\alpha$, the supremum of which corresponds to $-\tau_{q}$
(see (\ref{eq:tau_q_typ_GOE}) for example). For $W=2.6$ the left
\emph{local} maximum at $\alpha\simeq0.1$ does not appear to flow
with system size, while the \emph{right} local maximum at $\alpha\simeq\alpha_{0}$
drifts upward. Nevertheless this upward flow is bounded from above
by the normalization conditions (\ref{eq:f_alpha_bounds}) (shown
by black dashed lines in the figure) and thus the right local maximum
at $\alpha\simeq\alpha_{0}$ cannot overcome the one at $\alpha\simeq0.1$
even in thermodynamic limit. While it is possible that there is a
very slow downward flow of the left local maximum, which will eventually
restore GOE, we do not see it within the available system sizes. Further
support for possible multifractality at $W=2.6$ can be obtained by
examining the well-known symmetry of multifractal spectrum, which
can be analytically derived for wavefunction envelopes in multifractal
states of various models \citep{Evers2008a}, 
\begin{equation}
f\left(\alpha\right)=f\left(2-\alpha\right)+\alpha-1.\label{eq:MF-symmetry}
\end{equation}
While this symmetry is not necessarily satisfied when multifractality
is present (for example it fails in localized phases and in some extended
phases with Poisson statistics), it serves as an additional indication
of multifractality.

In the insets of Fig.~\ref{fig:falpha-qalpha} we test this symmetry
for the maximal available system size $L=24$. To suppress the effects
of zeros of the eigenstates we only examine the symmetry in the regime
where the tail of $f\left(\alpha\right)$ is significantly above its
ergodic value $\left(3-\alpha\right)/2$ (see Eq.~(\ref{eq:f_alpha_GOE})),
which for our data occurs for $W\geq2.6$ (see Fig.~\ref{fig:falpha}).
In this range of disorder strengths the multifractal symmetry~(\ref{eq:MF-symmetry})
is satisfied (insets of Fig.~\ref{fig:falpha-qalpha}), while in
the complementary range, $W\leq2.2$ the symmetry doesn't apply. This
indicates a possible multifractal phase for $W>2.2$.

\section{Summary and discussion}

In this work we have conducted a detailed large-scale numerical study
of multifractal properties of eigenstates on the delocalized side
of the many-body localization transition (MBL). This phase is known
to have a number of anomalous dynamical features, such as subdiffusive
transport, sublinear entanglement entropy growth and suppressed spreading
of information \citep{Luitz2016c}. For the single-particle case,
suppressed relaxation and dynamics are often associated with spatial
sparseness of the underlying eigenstates \citep{Ketzmerick1997,Ohtsuki1997}.
A natural question to ask is whether a similar relation exists also
in the many-body case, namely if sparseness of the eigenstates in
Hilbert space implies slow relaxation and suppressed transport of
\emph{local} observables. In this work we answer this question in
the \emph{negative}, by identifying a large fraction of the delocalized
phase which is consistent with ergodicity, while still showing a clear
signature of subdiffusion and slow relaxation in both numerical and
experimental data. We reach this conclusion by a careful analysis
of the finite size flows of eigenstate coefficient distributions,
moments of these distributions and their spectrum of multifractal
dimensions. Our analysis focuses on the computational basis, where
the basis states are labeled by the eigenvalues of the local $\hat{S}_{i}^{z}$
operators. This is the natural basis for the XXZ chain in the context
of MBL since it is compatible with the disorder and is naturally linked
to the hopping problem in Hilbert space. To the best of our knowledge,
the multifractal spectrum of the disordered XXZ chain has not been
studied before due to severe finite-size behavior and non-monotonic
behavior (see Figs.~\ref{fig:iq_vs_iq_goe} and \ref{fig:falpha}
for example), which hindered reliable extrapolation to the thermodynamic
limit.

In this work we focus on standard multifractal probes, namely, on
the spectrum of fractal dimensions $f(\alpha)$ and on its Legendre
transform, the critical exponent $\tau_{q}$ of the generalized IPR.
We distinguish between mean and typical averaging of $\tau_{q}$ over
different eigenstates and disorder strengths. All the measures we
study provide a coherent picture of a steady flow towards the predictions
of GOE for disorder strengths $W\lesssim2.6$. The average $\tau_{q}$
deviates from the ergodic limit $q-1$ only in the atypical region,
$f(\alpha)<0$, which corresponds to bin counts decreasing with the
system size in the wavefunction histogram. At the same time typical
$\tau_{q}$ and multifractal spectrum $f(\alpha)$ demonstrate a district
flow towards GOE values. The typical $\tau_{q}$ show a slight non-concavity,
in this disorder interval, due to sub-leading non-linear finite-size
effects similarly to the behavior in Anderson localization \citep{Evers2008a,Mildenberger2002},
however this non-concavity is within the error bars and can thus can
be safely ignored. A slight discrepancy compared to the GOE prediction,
is observed for $W<1$, as an excess of small values of the eigenstate
coefficients compared to GOE. Although it is consistent with a so-called
weak ergodic phase, where the wavefunction occupies a finite, but
tiny fraction of the Hilbert space, observed in several single-particle
and many-body systems \citep{Baecker2019,Nosov2019correlation,BogomolnyPLRBM2018},
we argue that this discrepancy is a result of proximity to an integrable
point at $W=0$, and should --- if this is indeed the case --- disappear
either in the thermodynamic limit, or if integrability is broken.
We leave the verification of this prediction to a future study.

For larger disorder strengths, our analysis becomes unreliable, due
to slowing down of finite-size flows. While we cannot rule out a slow
residual flow to GOE (which would provide an alternative explanation
in line with strong finite-size effects \citep{Panda2019,Abanin2019a,Sierant2019a,Weiner2019a,Khemani2017}),
we don't observe it within our range of accessible system sizes .
At this disorder strength, both average and typical $\tau_{q}$ deviate
from their ergodic limit $q-1$ at $q\gtrsim1$, which is consistent
with the saturation of the down-flow of $f(\alpha)$ at $\alpha\lesssim1$.
Moreover, for $W=2.6$ the multifractal spectrum $f(\alpha)$ perfectly
satisfies one of its basic symmetries, which would be consistent with
multifractality of the eigenstates in this region. Given the immense
numerical cost of our calculations, we could only compute the spectrum
in a limited range of disorder strengths across the delocalized phase.
Combined with the slow finite size flows at stronger disorder, we
cannot determine whether the region consistent with multifractality
shrinks to the critical point when the system size is increased, as
was claimed in Ref.~\citep{Serbyn2016a}. It would be interesting
to study this important question in more detail in the future.

One of the central outcomes of our study suggests that the previously
observed anomalous dynamics is \emph{not} related to multifractality
of many-body eigenstates. However ,since multifractal features are
generically basis dependent, one can wonder whether the outcome of
our study changes with the change of the basis. While it is difficult
to predict the effect of a basis rotation without performing actual
calculations, it is clear that for GOE eigenstates the multifractal
features (i.e. non-fractal in this case) do not depend on the basis,
since the GOE distribution is invariant under orthogonal transformation
\citep{mehta}. However, this is only true for \emph{almost all} bases
(for example the eigenstate basis is clearly not a good basis to study
multifractality). In contrast, for truly multifractal states both
in single-particle and many-body systems multifractal and localization
properties are drastically basis-dependent. In the Anderson localization
community the spatial basis presents a natural choice where the localization
transition also show changes in the level statistics \footnote{Recent developments show that in correlated models the spectrum properties
are related to localization in several bases (cf. Ref.~\citep{Nosov2019correlation}).}, but there is no such obvious choice for the many-body case. One
good candidate for such a basis, which can be directly tied to relaxation
of local observables, is the family of bases generated by \emph{locally}
exciting the eigenstates of the system \citep{Serbyn2016a}.
\begin{acknowledgments}
The authors acknowledge fruitful discussions with Nicolas Macé, Fabien
Alet and Nicolas Laflorencie. This research was supported by the Israel
Science Foundation (grants No. 527/19 and 218/19). We acknowledge
PRACE for awarding access to HLRS's Hazel Hen computer based in Stuttgart,
Germany under grant number 2016153659. Our code is based on the PETSC
\citep{petsc-efficient,petsc-user-ref}, SLEPC \citep{slepc-toms,slepc-manual}
and Strumpack \citep{ghysels_efficient_2016,ghysels_robust_2017}
libraries. I. M. K. acknowledges the support of German Research Foundation
(DFG) Grant No. KH 425/1-1 and the Russian Foundation for Basic Research
Grant No. 17-52-12044.
\end{acknowledgments}

\appendix

\section{Matrix elements}

\begin{figure}[h]
\includegraphics[width=0.9\textwidth]{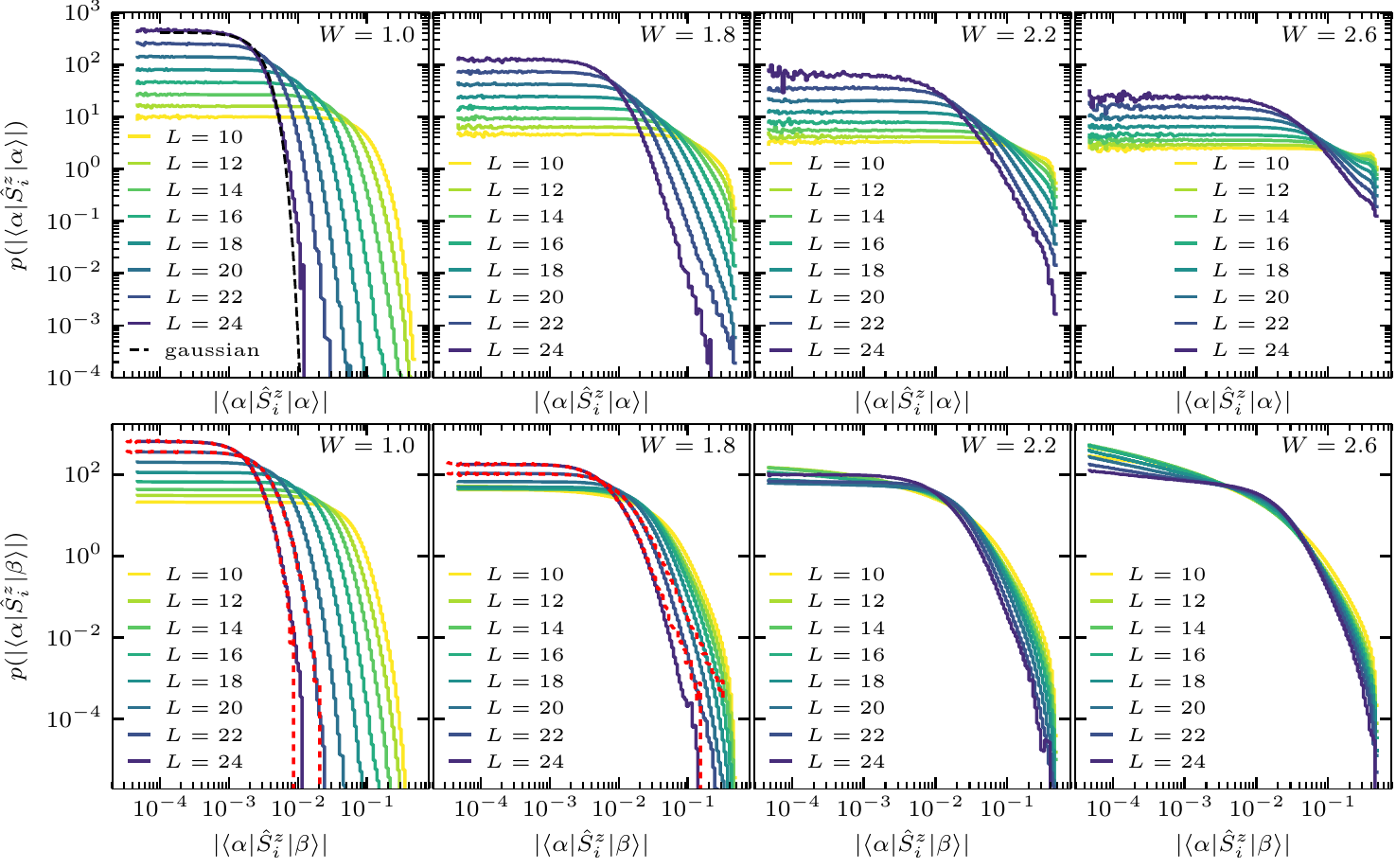} \caption{Distribution of diagonal (top) and offdiagonal (bottom) matrix elements
of the local $\hat{S}_{i}^{z}$ operator as a function of disorder
strength $W$ in the eigenbasis $\left\{ \protect\ket{\alpha}\right\} $
of the Hamiltonian. For each disorder realization, $50$ eigenstates
closest to middle of the many-body spectrum. For each system size
and disorder strength $10^{2}\dots10^{4}$ disorder realizations are
included, as well as all positions $i$ in the chain. Note that for
the diagonal matrix elements $\protect\bra{\alpha}\hat{S}_{i}^{z}\protect\ket{\alpha}$
the distribution for each disorder realization has a (slightly) nonzero
mean, which we subtracted here (cf. discussion in Ref.~\citep{colmenarez_statistics_2019}).
The red dashed histograms in the bottom row for $W=1.0$ and $W=1.8$
correspond to the (rescaled) distribution of the diagonal matrix elements
$\sqrt{2}p\left(\protect\bra{\alpha}\hat{S}_{i}^{z}\protect\ket{\alpha}/\sqrt{2}\right)$
for comparison. For stronger disorder, the distributions of diagonal
and offdiagonal matrix elements are so strikingly different that we
do not show them in the same panel here.}
\label{fig:matel-dist}
\end{figure}

In this appendix, we turn our attention to the analysis of the distributions
of matrix elements of the local magnetization $\hat{S}_{i}^{z}$ in
the eigenbasis of the Hamiltonian. Similarly to our analysis of the
eigenstates coefficients in the main text, for each disorder realization
we consider 50 eigenstates $\ket{\alpha}$ of the Hamiltonian $\hat{H}$
with an eigenvalue $E_{\alpha}$ closest to middle of the many-body
spectrum $\left(E_{\text{max}}-E_{\text{min}}\right)/2$. These eigenstates
correspond roughly to infinite temperature. We note however, that
since we study the microcanonical ensemble with $S_{z}^{\text{tot}}=0$
here, where $\tr_{S_{z}=0}H\neq0$, in each disorder realization there
is a slightly different ``effective temperature'', which we correct
by subtracting the mean of the diagonal matrix elements (computed
over the extracted eigenstates) for each disorder realization (cf.
discussion in Ref. \citep{colmenarez_statistics_2019} and in particular
Appendix B therein).

We complement our previous work in Refs.~\citep{Luitz2016,Luitz2016b},
by calculating the distribution of the matrix elements $\bra{\alpha}\hat{S}_{i}^{z}\ket{\beta}$
of the local $\hat{S}_{i}^{z}$ in the eigenbasis $\left\{ \ket{\alpha}\right\} $
of the Hamiltonian, with a massively improved statistics and one additional
system size $(L=24)$. We also add logarithmic binning of the histograms,
a direct distribution of diagonal matrix elements rather than their
differences as well as a direct comparison of diagonal $\alpha=\beta$
and offdiagonal $\alpha\neq\beta$ matrix elements distributions.

Fig.~\ref{fig:matel-dist} shows the results for the logarithmically
binned probability density of diagonal $\left|\bra{\alpha}\hat{S}_{i}^{z}\ket{\alpha}\right|$
and offdiagonal $\left|\bra{\alpha}\hat{S}_{i}^{z}\ket{\alpha}\right||$
matrix elements of $\hat{S}_{i}^{z}$ for disorder strengths $W=1.0,\,1.8,\,2.2,\,2.6$,
which are well on the delocalized side of the phase diagram, for system
sizes $L=10,12\dots,22,24$. The logarithmic binning highlights the
maximum of the distributions, where the matrix elements are closeset
to zero. At weak disorder $W\lesssim1.0$ we note that both diagonal
and offdiagonal matrix elements assume a distribution very close to
Gaussian as predicted by ETH \citep{Srednicki1994}. Furthermore,
the prediction from random matrix theory that distributions of diagonal
and offdiagonal matrix elements should be directly related \citep{DAlessio2015,Foini2019}
is verified to very high precision (dashed red lines in the lower
panels of Fig. \ref{fig:matel-dist} are diagonal distributions for
$L=22,24$ (renormalized by $\sqrt{2}$ in order to take account of
convolution of two gaussian distributions) in comparison to offdiagonal
distributions shown in color). It is clear however (as was shown in
Ref.~\citep{Foini2019}), that for stronger disorder $W\gtrsim1.8$,
this correspondence is violated.

For the diagonal matrix elements the shape of the maximum appears
to be Gaussian (flat on a logarithmic scale), however the tails of
the distribution deviate from Gaussian distributions at disorder strengths
$W\gtrsim1.0$. The double logarithmic scale reveals a long straight
tail, particularly well developed for $W=1.8$ and $W=2.2$, which
seems to be consistent with a power law tail over more then one decade.For
the offdiagonal matrix elements the tail seems to decay faster than
a power law. In contrast to the diagonal matrix elements, there is
a significant excess weight at small values of the offdiagonal matrix
elements $\left|\bra{\alpha}\hat{S}_{i}^{z}\ket{\beta}\right|$, which
seems to scale to zero for $W=2.2$ but survives up to at least $L=24$
for $W=2.6$. The scaling of the matrix element distribution variance
inversely with Hilbert space dimension is well visible at weak disorder
$W=1.0$ in the (almost) equidistant distributions for both diagonal
and offdiagonal matrix elements and was analyzed in detail in Refs.~\citep{Luitz2016,Luitz2016b}.
Increasingly strong deviations from this scaling are observed at stronger
disorder, which were connected to subdiffusive transport \citep{Luitz2016b}.

\bibliographystyle{apsrev4-1}
\bibliography{lib_yevgeny,lib_manual}

\end{document}